\documentclass[12pt]{article}
\usepackage{cjp}

\begin{document}
\def\Journal#1#2#3#4{{#1} {\bf #2}, #3 (#4)}

\def\NCA{\em Nuovo Cimento}
\def\NIM{\em Nucl. Instrum. Methods}
\def\NIMA{{\em Nucl. Instrum. Methods} A}
\def\NPB{{\em Nucl. Phys.} B}
\def\PLB{{\em Phys. Lett.}  B}
\def\PRL{\em Phys. Rev. Lett.}
\def\PRD{{\em Phys. Rev.} D}
\def\PRP{{\em Phys. Rept.}}
\def\ZPC{{\em Z. Phys.} C}
\def\JMP{\em J. Math. Phys.}
\def\CMP{\em Commun. Math. Phys.}

\def\ep{\epsilon}
\def\vep{\varepsilon}
\def\be{\begin{equation}}
\def\ee{\end{equation}}
\def\bea{\begin{eqnarray}}
\def\eea{\end{eqnarray}}
\def\sla{\raise.15ex\hbox{$/$}\kern-.57em}

\title
{\vspace{-5.0cm}
\begin{flushright}{\normalsize RUHN-99-5}\\
\end{flushright}
\vspace*{2.5cm}
Overlap\footnote{Talk at Chiral '99, Sept 13-18, 1999, Taipei, Taiwan.}
}
\author{H. Neuberger}
\address{Department of Physics and Astronomy\\
	Rutgers University\\
	Piscataway, NJ 08855, USA.}
\maketitle
\begin{abstract}
The overlap formulation of regulated vectorial
and chiral gauge theories is reviewed. Ostensibly 
new constructions, based on the Ginsparg-Wilson 
relation are essentially just overlap with new notation. 
At present there exists no satisfactory realization of chiral
symmetries outside perturbation theory which
is structurally different from the overlap. 
\end{abstract}
\section{Introduction}
My talk is divided into five parts. 
In the first part the relevance
of the chiral fermion issue to fundamental particle physics and to
numerical QCD will be explained. In the second part the basic ideas
of the overlap construction will be reviewed. In the third
part I shall focus on the algebraic (as opposed to kinematic)
meaning of the overlap. In the fourth part I shall discuss in
greater detail the overlap Dirac operator on the lattice. In
the last part I shall deal with anomalies, the associated Berry
phases and F{\" o}rster-Nielsen-Ninomiya gauge averaging in the context
of naturalness. My main partner in the overlap 
construction was R. Narayanan. I
have also collaborated with P. Huet, Y. Kikukawa, A. Yamada 
and P. Vranas.
Important contributions to the overlap development were also made by
S. Randjbar-Daemi and J. Strathdee.

\section{Relevance}
Fermion chirality is one of the most fundamental properties 
of Nature where it appears in association with gauge interactions.
This combination can decouple from a more fundamental theory
if anomalies cancel. It is however not trivial to achieve
this decoupling outside perturbation theory, and for
a while it was even thought an impossible feat. The overlap
provides an example where the decoupling works in a non-perturbative
framework. In a unitary framework the mechanism requires an 
infinite number of fermion
degrees of freedom per unit four space-time volume. It would
be a major achievement for lattice field theory if it turned
out that a similar mechanism is operative in Nature. More details
regarding general relevance can be found in \cite{pisa}.

\section{Basic structure}

The overlap is an outgrowth of a long sequence of papers
starting from the fundamental discoveries in \cite{basic},
\cite{stora_zumino}, \cite{reviews} and more specifically
\cite{callan_harvey}, \cite{boya}, \cite{kaplan}, \cite{frolov_slavnov}.
My first papers on the subject were all written with Narayanan
who stayed my collaborator for a long time \cite{neub_nar_plb1},
\cite{neub_nar_npb1,neub_nar_npblong,twod_chiral}. In parallel, 
important contributions to the overlap were made in \cite{seif}.
Later, I continued to simplify the overlap in the vector-like
context \cite{neub_plb_massless}. 

Let us start with a formally vector-like gauge theory:
\begin{equation}
{\cal L}_\psi =\bar\psi \sla D \psi + \bar \psi (P_L {\cal M} +
P_R {\cal M}^\dagger )\psi .
\end{equation}

$\bar\psi$ and $\psi$ are Dirac fermions and the mass matrix ${\cal M}$ is
infinite. It has a single zero mode but its adjoint has no zero modes.
As long as ${\cal M}{\cal M}^\dagger > 0$ this setup 
is stable under small deformations of the mass matrix implying
that radiative corrections will not wash the zero mode away. 

Kaplan's domain wall suggests the following realization:
\begin{equation}
{\cal M} = -\partial_s - f(s), 
\end{equation}
where $s\in (-\infty , \infty )$ and $f$ is fixed at $-\Lambda^\prime$
for negative $s$ and at $\Lambda$ for positive $s$ ($\Lambda^\prime ,
\Lambda > 0$. There is no
mathematical difficulty associated with the discontinuity at $s=0$.

The infinite path integral over the 
fermions is easily ``done'': on the positive
and negative segments of the real line respectively one has propagation with an
$s$-independent ``Hamiltonian''. The infinite extent means that at $s=0$ 
the path integrals produce 
the overlap (inner product) between the two ground states of the
many fermion systems corresponding to each side of the origin in $s$.
The infinite extent also means infinite exponents linearly proportional
to the respective energies - these factors are subtracted. One is
left with the overlap formula which expresses the chiral determinant
as $\langle v^\prime \{ U \} | v \{ U \} \rangle$. The states
are in second quantized formalism. By convention, they
are normalized, but their phases are left arbitrary. This
ambiguity is essential, as we shall see later on. It has no
effect in the vector-like case. 
In first quantized formalism the overlap is:
\begin{equation}
\langle v^\prime \{ U \} | v \{ U \} \rangle = 
\det _{k^\prime k} M_{k^\prime k} .
\end{equation}
The elements of the matrix $M$ are 
the overlaps between single body wave-functions, $M_{k^\prime k}=
v^{\prime \dagger}_{k^\prime} v_k$. 
The $v^\prime$'s span the negative
energy subspace of $H^\prime \sim \gamma_5 ( \sla D_4 +\Lambda^\prime )$
and the $v$'s span the negative
energy subspace of $H \sim \gamma_5 ( \sla D_4 -\Lambda)$. 
I used continuum like notation to emphasize that the Hamiltonians
are arbitrary regularizations of massive four dimensional Dirac
operators with large masses of opposite signs. 

The Hamiltonians only enter as defining the Dirac
seas and there is no distinction between the different levels within 
each sea; all that matters is whether a certain single particle state
has negative or positive energy. Thus, all the required information is
also contained in the operators $\epsilon =\varepsilon (H)$ and
$\epsilon^\prime = \varepsilon (H^\prime ) $ where $\varepsilon$ is the
sign function. Thus the $v^\prime$'s are all the $-1$ eigenstates of
$\epsilon^\prime$ and the $v$'s are all the $-1$ eigenstates of $\epsilon$.
To switch chiralities one only has to switch the sign of the Hamiltonians.
This is a result of charge conjugation combined with a particle-hole
transformation.

When $\Lambda^\prime$ is taken to
infinity in lattice units one is left with $\epsilon^\prime =\gamma_5$
with no gauge field dependence. 
On the other hand, $\epsilon$ always maintains
a dependence on the gauge background and its trace
gives the topological charge of the gauge field. 

More recently, a construction starting from the GW
relation \cite{ginsp_wils} has been presented as new \cite{luscher}. 
This is misleading, since the ``new'' construction 
merely reproduces the structure of the overlap.
The only differences are in technicalities 
surrounding the phase choice of the overlap.  
These technicalities are not trivial, but I would be surprised
if any new particle physics were produced by this effort. To be
sure, a successful completion of the phase choice program 
would be news in mathematical physics, and would make me happy.  
The proposal of the ``new'' construction was preceded by the observation
that the overlap satisfies the GW relation \cite{neub_prd}, \cite{neub_plb_gw} 
and that the overlap-GW relation is essentially one to one
\cite{narayanan}, \cite{chiu-zenkin}. Thus, it should have been obvious
that starting from the GW relation will lead to the overlap structure in
the chiral case. The insistence of the proponents of the 
``new'' construction to obscure its structural equivalence to
the overlap has generated much confusion and waste of time, deflecting
energy from some valid physics problems that are still open. For example, 
it is an important open question whether a genuinely different natural 
construction of chiral gauge theories is possible.

\section{Algebraic meaning}

The GW relation is a complicated way to write down what is
best described as the algebraic content of the overlap. It
is much simpler to rephrase and slightly generalize: The setup
is that of a ``Kato pair'' \cite{kato}, an arrangement further developed
in the Quantum Hall Effect 
context by Avron, Seiler and Simon \cite{ass} and others \cite{silv}.

The core objects are the two hermitian reflections,
$\epsilon$ and $\epsilon^\prime$ introduced above.
They are equivalent to two
orthogonal projectors $P,Q$, the Kato pair
($\epsilon=1-2Q,~\epsilon^\prime =1-2P$). 
Defining
\be
h={1\over 2}({\epsilon +\epsilon^\prime}),~~~s={1\over 2}({\epsilon
-\epsilon^\prime}),
\ee
we arrive at the following fundamental relations:
\be
h^2 +s^2 =1,~~~\{h , s\} =0 .
\ee
These relations allow us to write any element in the algebra
generated by $h$ and $s$ as $f_1 (h) + s f_2 (h^2) + hs f_3 (h^2)$,
where $h^2$ is a central element. Geometrically, $P$ and $Q$
generate two orthogonal decompositions of the entire space ${\cal V}$.
A natural operator mapping one subspace of one decomposition
to another subspace of the other decomposition was introduced
by Kato. It coincides with overlap Dirac operator.
\be
D_o ={1\over 2} (1+\epsilon^\prime \epsilon )=\epsilon^\prime h =
h \epsilon .
\ee
That it connects the two decompositions is evident from $PD_o = D_o Q$.
In the overlap context $D_o$ was found via the following 
easily proven identity: 
\be
det D_0 = |det M|^2 .
\ee

The following decomposition of the Hilbert space into
subspaces invariant under $h$ and $s$ holds:
\be
{\cal V} = Ker (h) \oplus Ker (s) \oplus {\cal V}_\perp .
\ee
$Tr h|_{{\cal V}_\perp} =Tr s|_{{\cal V}_\perp}=0$ by 
virtue of $\{h, s\}=0$. One also has
\be
Ker(h)=Ker(s-1)\oplus Ker (s+1),~~~~~Ker(s)=Ker(h-1)\oplus Ker (h+1) .
\ee

A special property of the Kato pair in the overlap context is
that $h,s$ are both gauge field dependent and that for all
backgrounds $Tr h = Tr s$. 
One can introduce now a definition of index
\be
{\rm index} (\epsilon^\prime ,\epsilon ) = dim Ker (s-1) - dim Ker (s+1)=
Tr s^{2n+1}
\ee
for any integer $n\geq 0$. To prove $n$-independence of the right hand
side use $s^2=1-h^2$ and $Tr h^{2k} s = Tr h^{2k-1}s h = - Tr h^{2k} s$.
Reversing the roles of $h$ and $s$ switches the sign of the index. 
It turns out that the index is exactly the 
topological charge of the gauge field as sensed by the 
overlap \cite{neub_nar_prl}. 
Moreover, if $s=1$ and $s=-1$
are isolated points of the spectrum of $s$ and both $Ker(s+1)$
and $Ker(s-1)$ are finite dimensional the definition extends to
$dim {\cal V}=\infty$.

Until now the roles of $h$ and $s$ were symmetrical, and there is
no massless fermion in sight.
The entire setup is pure mathematics. 
The kinematic part of the construction is in the relationship
between the reflections (or projectors) and the massive
Dirac operator. This is the basis of the lattice realization of
the overlap idea to be covered in the next section. Here, we shall
devote a few lines to explain why the Kato pair is an algebraic
set-up which is essentially equivalent to the Ginsparg Wilson
relation. The Ginsparg Wilson relation is:
\be
\{D, \gamma_5 \} = 2 D\gamma_5 R D .
\ee
One adds to the above the following requirements: 
(1) $\gamma_5$-hermiticity, $\gamma_5 D \gamma_5 = D^\dagger$;
(2) $[R,\gamma_\mu ]=0$; (3)  $R^\dagger =R > 0$; $R$ is local in the site
index. It goes without saying that $D$ and $R$ transform covariantly
under gauge transformations and that $D$ should be local in some sense.

Choosing $\epsilon^\prime=\gamma_5$ it is obvious that $D_o$
satisfies the GW relation
and the extra requirements with $R=1$. 
$\epsilon$, being associated with very
massive fermions, will be essentially local. In the lattice
context, as we shall see in the next section, some gauge backgrounds
need to be excluded if locality is to hold absolutely.

What may be less obvious is the other direction of the equivalence, 
namely, that the GW relation implies a Kato pair. 
Starting from the GW relation takes one back to the overlap situation
at the algebraic level, i.e., before the explicit form of the reflection
$\epsilon$ is given. The reflection $\epsilon^\prime$, which is quite free
in the overlap, is restricted in the GW case to $\gamma_5$. 

The first step is to eliminate $R$, by showing that any solution
with $R=1$ produces a solution with an arbitrary $R$. This can be done
in at least two ways \cite{chiu-new}. The first starts from
the main observation that the basic GW relation \cite{chiu-zenkin} 
can be written as
\be
\{D^{-1}(R) - R,\gamma_5 \}=0
\ee
and therefore setting $D^{-1}(R) - R=D^{-1}(1) - 1$ ensures that 
$D(R)$ is a solution if $D(1)$ is. Although the inverse
of $D_\chi \equiv {1\over  {D^{-1}(R) - R}}$
is non-local by the well known no-go 
theorems, $D^{-1}_\chi$  has only
zeros at the ``doubler'' locations in the free case, and
$R$ removes these zeros from $D^{-1}(R)$, 
so that $D(R)$ ends up local. Once $D(R)$ is local and doubler free 
in the free case, locality in the presence of lattice backgrounds 
that are sufficiently close to the continuum is assured. 
The second way is to set $D(R)={1 \over \sqrt{R}} D(1) 
{1\over \sqrt{R}}$.
So, nothing is lost or gained in principle by fixing $R=1$. 

For $R=1$ it is trivial to go back to the reflections: one just
looks at the formula for $D_o$ to define the operator
$\gamma_5 [2D(1)-1]$. It is trivial that this operator is hermitian
and squares to unity by the basic GW formula \cite{narayanan}. 
So, $D(1)$ is
just another $D_o$. Now it is trivial to extract $h$ and $s$,
since $\epsilon^\prime$ is given as $\gamma_5$. Also, it is a trivial
matter to go to the matrix $M$ and the chiral case.

\section{Concrete realization}

The massless fermions enter as a representation
of the difference between Dirac operators with masses of
opposite sign. There are no particular 
difficulties to regularize {\it massive}
Dirac fermions, in the continuum or on the lattice. 
On the lattice the spectra of the Dirac
operators get compactified and the regularization is fully
non-perturbative.

First, some notation:
Let $U_\mu (x)$ denote  $SU(n)$ link matrices on a finite 
$d$-dimensional hypercubic lattice.
The directional parallel transporters $T_\mu$
are:
\be
T_\mu (\psi ) (x) = U_\mu (x) \psi (x+\hat\mu) .
\ee
Let $G(g)$ with $(G(g)\psi)(x) =g(x)\psi(x)$ describe
a gauge transformation. 
\be
G(g) T_\mu (U) G^\dagger (g) = T_\mu (U^g ) ,~~
{\rm where,}~~
U^g_\mu (x) = g(x) U_\mu (x) g^\dagger (x+\hat\mu ) .
\ee

A lattice replacement of the massive continuum Dirac operator, $D(m)$,
is an element in the algebra generated 
by $T_\mu,~ T_\mu^\dagger ,~ \gamma_\mu$.
Thus, $D(m)$ is gauge covariant. 
The Wilson Dirac operator,
$D_W (m)$ can be written as:
\be
D_W = m+\sum_\mu(1- V_\mu );~~~~V_\mu^\dagger V_\mu =1;~~~V_\mu=
{{1-\gamma_\mu}\over 2} T_\mu +{{1+\gamma_\mu}\over 2} T_\mu^\dagger .
\ee
The hermitian 
Wilson Dirac operator is $H_W (m)=\gamma_{d+1} D_W (m) $. In terms
of the continuum covariant derivative $D^c_\mu$, for lattice spacing $a$,
$V_\mu = e^{-a\gamma_\mu D^c_\mu}$, with no sum on $\mu$. 

We choose the following norm definition for matrices $A$: $\| A\|=
\sqrt{\lambda_{\rm max} (A^\dagger A ) }$.
The norm of a gauge covariant matrix is gauge 
invariant. $\lambda_{\rm max}$ is a maximal eigenvalue. 

Our two reflections are defined by:
\be
\epsilon^\prime = \lim_{m\to +\infty} \varepsilon(H_W (m))
=\gamma_{d+1};~~~~
\epsilon =\varepsilon ( {H_W (m)})
~~{\rm with}~~-2<m<0 .
\ee
For $\epsilon$ to be defined everywhere we need $H_W^2 (m) > 0$.
This can be assured if the pure gauge action enforces an upper bound
on the norms of $[T_\mu , T_\nu ]$ \cite{neub_bounds}. 
The following inequality, when meaningful, 
specifies what constraint would be sufficient:
\be
\left [ \lambda_{\rm min} (H_W^2 (m ))\right ]^{1\over 2}\geq
\left [ 1 - (2+\sqrt{2}) 
\sum_{\mu > \nu }\|[T_\mu , T_\nu ] \| \right ]^{1\over 2} -|1+m|, ~~~~
-2 < m < 0 .
\ee
The upper bound on $\|[T_\mu , T_\nu ] \| = \|1 - P_{\mu\nu}\|$ 
is compatible
with the continuum limit, where the $T_\mu$'s almost commute: 
$[T_\mu , T_\nu ]^\dagger [T_\mu , T_\nu ]= (1-P_{\mu\nu})^\dagger
(1-P_{\mu\nu})$, $P_{\mu\nu} = T_\nu^\dagger T_\mu^\dagger T_\nu T_\mu$,
$(P_{\mu\nu}\psi)(x) = U_{\mu\nu} (x) \psi (x)$,
$U_{\mu\nu}(x)=
U_\nu^\dagger (x) U_\mu^\dagger (x+\hat\nu ) U_\nu (x+\hat\mu ) U_\mu (x)$.
Any pure gauge action with the right continuum limit and close to it
will strongly
prefer configurations where all $U_{\mu\nu}(x)$ are close to unit matrix.

On the other hand, a latticized instanton can be smoothly deformed
to a trivial configuration, so there exist gauge field backgrounds
for which $H^2_W \{ U \} (m) $ has an exact zero mode. This
is true of any $m$ in the range $(-2,0)$. The above analysis tells
us that at least one plaquette will be relatively big for
this configuration. Unless we enforce the constraint fully,
there always will be possibly rare configurations 
where $\epsilon$
is not well defined. 

To fully appreciate the differentiation between the algebraic character
of the GW relation and the more complete framework of the overlap, let me
give some examples of bad solutions to the GW relation:
\bea
D^{(1)}_{GW} = {1\over 2} (1+\gamma_5 \varepsilon (H_W (m_1 )),&~~
D^{(2)}_{GW} = {1\over 2} (1-\gamma_5 \varepsilon (H_W (m_2 )); 
~~m_{1,2} > 0 .\\
D^{(3)}_{GW} = {1\over 2} (1-\gamma_5 \varepsilon (H_W (m_3 ));&
~~-2< m_3 <0 .\eea
Solutions (1) and (2)
have always trivial index, (1) has no massless fermions at all,
while (2) has 16. (3) is motivated by the $h,s$ symmetry, describes 15
fermions and has a nontrivial index, but 7 fermions contribute
to it one way, and 8 the other.

\section{Anomalies}
The absolute value of $\langle v^\prime | v \{ U\} \rangle $ is well
defined and gauge invariant, but the complex number itself
is defined only up to phase. In the general case we have a
$U(1)$ bundle defined over the space of gauge fields with
$H_W^2 \{ U \} (m) > 0$, a gauge invariant condition. $|v^\prime \rangle$
is gauge field independent and gauge invariant
and $|v\{ U\} \rangle$ comes from a gauge
covariant Hamiltonian. Thus, the $U(1)$ bundle can be viewed
as defined over the space of gauge orbits. With the exclusion
of the points $det H_W \{ U \} (m) = 0$ this space can support non-trivial
$U(1)$ bundles as required for anomalies. 

A more familiar way
for physicists is to think in terms of Berry phases. The main
point is that the matrices $H_W \{ U \}$ are smooth in the link variables
and this smoothness is inherited by the states $|v\{U \} \rangle$. With
it come the inevitable Berry phase factors. The Berry phase factors
come from parallel transport on the space of gauge fields with
Berry's connection ${\cal A}$. Assuming some choice of states
$|v \{ U \} \rangle$ (possibly requiring patches with overlays)
Berry's connection is the following one-form over each patch
in the space of gauge fields:
\be
{\cal A} = \langle v \{ U \} |d v \{ U \} \rangle .
\ee
Berry's phases come from Wilson loop factors associated with
${\cal A}$. This makes it obvious why they are independent of the
choice of phases for the $|v \{ U \} \rangle$. The non-triviality
of Berry's phase is locally measured by the associated abelian field
strength, a globally defined two-form that usually does not vanish:
\be
{\cal F} = d{\cal A} .
\ee
While ${\cal A}$ depends on the phase choices of the
$|v \{ U \} \rangle$, ${\cal F}$ does not, a fact which
is reflected by expressing ${\cal F}$ in terms of the reflection
$\epsilon =\varepsilon (H_W (m))$ alone:
\be
{\cal F}=-{1\over 4} Tr [\varepsilon (H_W (m))~d\varepsilon (H_W (m)) ~
d\varepsilon (H_W (m)) ] .
\ee

To understand the meaning of ${\cal A}$ let us review briefly
one way of looking at anomalies in the continuum. We are
restricting our attention to the non-abelian case. After all,
in the four dimensional 
abelian case the gauge coupling isn't asymptotically free
and the entire issue of mathematical existence of an interacting chiral
gauge theory is a moot point. 

Once one has some definition of a regulated chiral determinant
one can take a variation with respect to the vector potential
$A_\mu (x)$ to get a current, a nonlocal functional 
$J_{\rm consistent}^\mu (A)$ of the gauge field $A_\mu (x)$.
If the regulated chiral determinant is gauge invariant
$J_{\rm consistent}^\mu (A)$ transforms covariantly under
gauge transformations, but when there are anomalies
and the regulated chiral determinant is not gauge invariant
$J_{\rm consistent}^\mu (A)$ has a complicated transformation.
However, one can always define a covariant current,
$J_{\rm covariant}^\mu (A)=J_{\rm consistent}^\mu (A)-\Delta J^\mu (A)$
which does transform covariantly. Moreover, $\Delta J^\mu (A)$ is
a local, calculable functional of $A_\mu (x)$ \cite{bardeen_zumino}. 
One cannot use
$J_{\rm covariant}^\mu (A)$ to reconstruct a gauge invariant regulated
determinant because it cannot be written
as the variation of a functional of $A_\mu (x)$ since it does not obey
the Wess-Zumion consistency conditions. 
If anomalies do cancel one can find a regularization
(at least in perturbation theory) with $\Delta J^\mu (A) =0$;
then $J_{\rm consistent}^\mu (A)=J_{\rm covariant}^\mu (A)$ can be regulated
first and a regulated gauge invariant determinant can be constructed. 

The main observation is that Berry's
connection ${\cal A}$ on the lattice (when summed over
all fermion multiplet contributions) plays the role of
$\Delta J^\mu (A)$ \cite{neub_geom}. 
To see this we view the lattice currents 
as one forms over the space of gauge fields.
\be
J_{\rm consistent}={{d\langle v^\prime | v \rangle }\over
{\langle v^\prime | v \rangle }}=
{{\langle v^\prime | d v \rangle_\perp }\over
{\langle v^\prime | v \rangle }} +{\cal A} .
\ee
By definition, $\langle v | d v \rangle_\perp =0$. The covariant
current is
\be
J_{\rm covariant}={{\langle v^\prime | d v \rangle_\perp }\over
{\langle v^\prime | v \rangle }} .
\ee
and it is quite obvious that it has to transform covariantly because
it is unaffected by a phase change in the choice of the
$|v  \{ U \} \rangle$ and hence is insulated of any gauge breaking
step.

${\cal A}$ only depends on the overlap of the state  $|v  \{ U \} \rangle$
with a state $|v  \{ U^\prime \} \rangle$ with $U^\prime$ close
to $U$. This is in contrast to the overlap or the currents
where $|v  \{ U \} \rangle$ is overlapped with a very different
state, $|v^\prime \rangle$. This is why ${\cal A}$ is local in
the gauge fields but the currents are not \cite{neub_geom}, \cite{daemilast}. 

This interpretation of ${\cal A}$ is made solid by the following
result: Consider an anomalous theory. One can find a two
parameter family of gauge backgrounds which becomes a two torus
when restricted to orbit space. ${\cal F}$ can also be  restricted to orbit
space because it is explicitly gauge invariant. One can integrate 
${\cal F}$ over this two torus and the result is proportional
to an integer. The integer is precisely the same as the one
entering the anomaly coefficient. The result can be easily shown to
be consistent with a known continuum expression. On the lattice
this result means that there are ``monopole'' like singularities
in ${\cal A}$. These singularities cannot be removed
by small deformations of the matrix $H_W \{ U \} (m)$. Thus, we
cannot envisage altering $H_W \{ U \} (m)$ so that ${\cal A}$
vanish. But, if anomalies do cancel the obstruction is removed.

On the basis of this I propose two conjectures \cite{neub_geom}:
\begin{itemize}
\item Iff anomalies cancel can one deform the total Hamiltonian so
that one attains ${\cal F}=0$ without the overlap going through
any singularity. 
\item If ${\cal F}=0$ there is a natural phase choice for
$|v  \{ U \} \rangle$, determined by parallel transport with
respect to an ${\cal A}$, such that the action of gauge 
transformations is non-projective: For any gauge transformation
$g(x)$, its representation on the fermions, $G(g)$ obeys
\be
G(g) |v  \{ U \} \rangle = |v  \{ U^g \} \rangle .
\ee
\end{itemize}
If these conjectures are true it would follow that one could
fine tune $H_W \{ U \} (m)$ so that, provided anomalies cancel,
a natural smooth gauge invariant phase choice for
the overlap exists. 

Still, this solution, if it indeed works, looks somewhat contrived
for Nature because it requires a fine tuning of the Hamiltonian used
in the construction. I believe that this fine tuning is not necessary.
If a fine tuned Hamiltonian exists, any Hamiltonian close enough to
it would also work, although a residual gauge dependence would be
present. One would simply average over the gauge degrees of freedom
(integrate them out) and this would have no effect on the continuum
limit \cite{FNN}. If anomalies do not cancel, the gauge dependence
is always large enough that gauge averaging induces new non-local terms
in the action. Indeed, the gauge dependence contains a lattice
version of a nontrivial Wess-Zumino term. 
When anomalies cancel but one still has a small amount 
of gauge breaking, gauge averaging only adds some extra gauge invariant
local terms to the action. This has been shown numerically in a
two dimensional chiral model \cite{twod_chiral}.

There exists an important special case where there are anomalies but
${\cal F}$ vanishes. It is in four dimensions with gauge group
$SU(2)$ and Weyl fermions in the $I={1\over 2}, {5\over 2},\dots$
representation. Then the matrix $H_W $ can be made real by a global base
choice. Still, Berry's phase factors can take the  values $\pm 1$ and
thus Witten's anomaly is reproduced \cite{neub_z2}. 
This again shows that in the overlap
all anomalies are encoded in Berry's phases.

Although the overlap at present does not provide a rigorous
construction for chiral gauge theories in the non-abelian 
case, it has passed the
boundary of physical plausibility, relegating the completion
of this construction to the branch of mathematical physics.
For Physics the main question is: Does there exist a truly
new way to regulate chiral symmetries, or is the way opened
by Kaplan and by Frolov and Slavnov and subsequently realized by 
the overlap unique in some sense ?

\section {Acknowledgments}

My research at Rutgers is partially supported by the DOE under grant
\# DE-FG05-96ER40559. I wish to express my appreciation of the
immense hospitality and great effort invested by the organizers
of Chiral 99 in Taipei. In particular I wish to thank Ting-Wai Chiu
for doing so much to produce an inspiring and enjoyable meeting.

\end{document}